\documentclass[aps,superscriptaddress,floats,showpacs,floatfix,twocolumn]{revtex4}
\usepackage{bm}
\usepackage{graphicx}
\usepackage{subfigure}
\usepackage{amssymb}
\usepackage{amsmath}
\usepackage{color}
\usepackage[a4paper,left=1.5cm, right=1.3cm, top=2cm,bottom=2cm]{geometry}
\begin{document}
\newcommand{\joerg}[1]{\textcolor{red}{#1}}
\newcommand{\ambrish}[1]{\textcolor{blue}{#1}}

\title{Transition to turbulence scaling in Rayleigh-B\'{e}nard convection}
\author{J\"org Schumacher}
\affiliation{Tandon School of Engineering, New York University, New York, NY 11201, USA}
\affiliation{Institut f\"ur Thermo- und Fluiddynamik, Technische Universit\"at Ilmenau, P.O.Box 100565, D-98684 Ilmenau, Germany}
\author{Ambrish Pandey}
\affiliation{Institut f\"ur Thermo- und Fluiddynamik, Technische Universit\"at Ilmenau, P.O.Box 100565, D-98684 Ilmenau, Germany}
\author{Victor Yakhot}
\affiliation{Department of Mechanical Engineering, Boston University, Boston, MA 02215, USA}
\author{Katepalli R. Sreenivasan}
\affiliation{Tandon School of Engineering, New York University, New York, NY 11201, USA}
\affiliation{Department of Physics and Courant Institute of Mathematical Sciences, New York University, New York, NY 10012, USA}

\date{\today}

\begin{abstract}
If a fluid flow is driven by a weak Gaussian random force, the nonlinearity in the Navier-Stokes equations is negligibly small and the resulting velocity field obeys Gaussian statistics. Nonlinear effects become important as the driving becomes stronger and a transition occurs to turbulence with anomalous scaling of velocity increments and derivatives. This process has been described by V. Yakhot and D. A. Donzis, Phys. Rev. Lett. {\bf 119}, 044501 (2017) for homogeneous and isotropic turbulence (HIT). In more realistic flows driven by complex physical phenomena, such as instabilities and nonlocal forces, the initial state itself, and the transition to turbulence from that initial state, are much more complex. In this paper, we discuss the Reynolds-number-dependence of moments of the kinetic energy dissipation rate of orders 2 and 3 obtained in the bulk of thermal convection in the Rayleigh-B\'enard system. The data are obtained from three-dimensional spectral element direct numerical simulations in a cell with square cross section and aspect ratio 25 by A. Pandey et al., Nat. Commun. {\bf 9}, 2118 (2018). Different Reynolds numbers $1\lesssim {\rm Re}_{\ell}\lesssim 1000$ which are based on the thickness of the bulk region $\ell$ and the corresponding root-mean-square velocity are obtained by varying the Prandtl number ${\rm Pr}$ from $0.005$ to 100 at a fixed Rayleigh number ${\rm Ra}=10^5$. A few specific features of the data agree with the theory but the normalized moments of the kinetic energy dissipation rate, ${\cal E}_n$, show a non-monotonic dependence for small Reynolds numbers before obeying the algebraic scaling prediction for the turbulent state. Implications and reasons for this behavior are discussed.
\end{abstract}
\maketitle

\section{Introduction}
The question of small-scale universality of turbulence is at the core of turbulence research since its beginnings \cite{Taylor1935,Kolmogorov1941,Frisch1994}. If universality exists, statistical moments must follow well-defined scaling laws with respect to length and time scales, or to essential parameters such as the Reynolds number ${\rm Re}$. Most studies which are dedicated to this subject aim at the highest possible Reynolds numbers in experiments \cite{Sreenivasan1997} or simulations \cite{Kaneda2009,Yeung2015} in order to achieve sufficiently large range of scales separating the large and small ones in the flow. A different option is to study the statistics of gradients of the turbulent fields which are always supported at the smallest scales, and whose statistical moments must follow well-defined laws with respect to parameters such as ${\rm Re}$. For homogeneous and isotropic turbulence (HIT), a phase transition (to be described momentarily) from Gaussian to non-Gaussian statistics of velocity derivative moments, thus a transition to multiscaling, has been demonstrated in \cite{Yakhot2006} and more recently in \cite{Yakhot2017}. 

If the ideas proposed for this transition in the statistical properties are to have some general validity, they have to find application in more complex flows, such as wall-bounded shear flows \cite{Waleffe1997,Eckhardt2007,Smits2013} or thermal convection flows \cite{Chilla2012} as well. In this paper, we test these theoretical ideas for Rayleigh-B\'enard convection (RBC). The mechanisms of production of turbulent kinetic energy in this flow are connected to life cycles of characteristic coherent structures of the thermal boundary layers \cite{Malkus1954,Shishkina2005,Zhou2007,Chilla2012,Schumacher2016}, so the details are bound to be more complex than in homogeneous and isotropic turbulence. In particular, we will study here the scaling of moments of the kinetic energy dissipation rate with respect to Reynolds number. 

Our RBC flows evolve in large-aspect ratio cells with values of $\Gamma=25$. In contrast to isotropic turbulence and wall-bounded flows, the Reynolds number Re is not a prescribed parameter, but is a derived quantity related to the turbulent momentum transfer in response to the applied temperature difference, and is related to the Rayleigh number ${\rm Ra}$; another property of importance for this flow is the Prandtl number ${\rm Pr}$, which is the ratio of the kinematic viscosity $\nu$ of the fluid to the temperature diffusivity $\kappa$. Here, a range of small to moderate Reynolds numbers is established by varying ${\rm Pr}$ over more than four orders of magnitude for a fixed Ra \cite{Pandey2018}. The lower the Prandtl number, the higher the Reynolds number \cite{Schumacher2015}. We focus our attention on the bulk of the flow away from the boundary layers at the heated bottom and cooled top plates of the RBC setup. 

The manuscript is organized as follows. In section II we provide a self-contained review of the foundations of the theory. Section III presents the numerical model and defines the essential parameters of the convection runs. Section IV reports our results and interpretation, and the last section summarizes the conclusions.  

\section{Scaling of moments of the kinetic energy dissipation rate}

\begin{figure*}
\centering
\includegraphics[scale = 0.13]{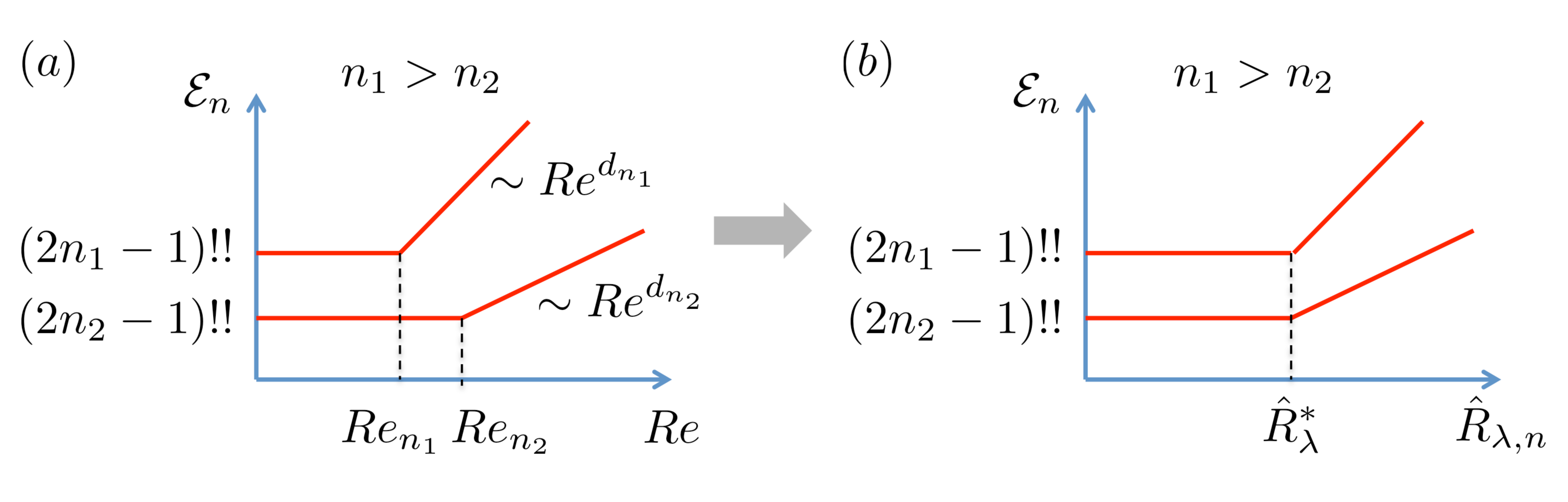}
\caption{Transition from Gaussian to non-Gaussian statistics for two normalized dissipation moments ${\cal E}_n$ in a homogeneous isotropic turbulent flow with moment orders $n_1>n_2$. (a) is the schematic when the large-scale Reynolds number Re is used, while (b) represents the situation when the Reynolds number is rescaled as described in the text.}
\label{Sketch}
\end{figure*}

Before describing the present work, it appears useful to recast the essential points of Yakhot and Donzis in a self-contained manner. Their analysis is specifically connected to the $x_1$-component of the velocity field $u_i(x_j,t)$ and the corresponding longitudinal derivative $\partial_1 u_1=\partial u_1/\partial x_1$. Throughout this work, we will use index notation, e.g. ${\bm x}=(x_1,x_2,x_3)=x_j$ in combination with the Einstein sum convection. The derivative moment of order $2n$ is given by
\begin{equation}
M_{2n} = \langle(\partial_1 u_1)^{2n}\rangle \quad\mbox{with}\quad M_{2n}=A_{2n} \frac{u_{\rm rms}^{2n}}{L^{2n}} {\rm Re}^{\rho_{2n}}\,.
\label{Mn}
\end{equation}
Here $u_{\rm rms}$ is the root-mean-square velocity obtained in practice from all three velocity components by a combined volume-time average which is assumed to be equal to an ensemble average $\langle\cdot\rangle$. The large-scale Reynolds number Re is given by ${\rm Re}=u_{\rm rms}L/\nu$, $\nu$ being the kinematic viscosity and $L$ the characteristic outer length scale. The prefactors $A_{2n}$ are dimensionless constants. The $n$-th order moment of the dissipation rate is given by
\begin{equation}
E_{n} = \langle\epsilon^n\rangle\quad\mbox{with}\quad E_{n}=B_{n}  \frac{u_{\rm rms}^{3n}}{L^{n}} {\rm Re}^{d_{n}}\,,
\label{En}
\end{equation}
where $B_n$ are dimensionless constants, and the dissipation rate field is given by
\begin{equation}
\epsilon({\bm x},t) =2\nu\, S^2({\bm x},t) \quad\text{with}\quad S^2=S_{ij}S_{ji} \,, 
\label{ediss}
\end{equation}
and $S_{ij}=(\partial_i u_j+\partial_j u_i)/2$ is the rate-of-strain tensor. It follows that the normalized moments of dissipation and longitudinal derivative are given by
\begin{align}
{\cal E}_{n} &= \frac{E_n}{(E_1)^n}=\frac{B_n}{B_1^n}{\rm Re}^{d_n-nd_1}\,,\\ 
{\cal M}_{2n}& =\frac{M_{2n}}{(M_2)^n}=\frac{A_{2n}}{A_2^n}{\rm Re}^{\rho_{2n}-n\rho_2}\,.
\end{align}
In a flow with Gaussian derivative statistics, one has normal scaling, i.e., $d_n=n d_1$ and $\rho_{2n}=n \rho_2$ leading to
\begin{equation}
{\cal M}_{2n} =\frac{A_{2n}}{A_2^n}=(2n-1)!!=\frac{B_n}{B_1^n}={\cal E}_n\,.
\end{equation}  
The double factorial is given by $(2n-1)!!=1\cdot 3\cdot 5\dots (2n-1)$. Beyond a critical Reynolds number ${\rm Re}^{\ast}\approx 100-200$, the velocity derivative moments follow algebraic scaling laws with respect to Re. The scaling exponents of the moments are then anomalous, that is, $d_n\ne n d_1$ and $\rho_{2n}\ne n\rho_2$. This transition depends on the order $n$ of the moment, i.e., the higher the order the smaller a ${\rm Re}_n^{\ast}$. The scaling exponents there can also be related to the anomalous scaling exponents, $\zeta_n$, for $n$-th order velocity increment moments in a fully developed inertial range of a high-Reynolds-number flow as shown in \cite{Yakhot2006}. These predictions were confirmed later in a high-resolution direct numerical simulations (DNS) \cite{Schumacher2007}. Normalized moments (of both ${\cal E}_n$ or ${\cal M}_n$) transition from Gaussian to non-Gaussian, and thence to turbulent regime at different Reynolds numbers. The situation is as described schematically in Fig.~\ref{Sketch}(a).  

In the spirit of Landau's theory of phase transitions, two ideas are now adapted: (i) The transition for all moment orders occurs at a unique and suitably redefined Reynolds number. The rescaling is partly familiar and uses, instead of Re, the Taylor microscale Reynolds number $R_{\lambda} =\sqrt{5/(3\langle\epsilon\rangle\nu)} \,u_{\rm rms}^2$. But this step alone is not enough; we redefine the microscale Reynolds number on the basis of a generalized velocity to be discussed below, in units of which the transition proceeds at a unique and order-independent Reynolds number, $\hat{R}_{\lambda,n}^{\ast}=R_{\lambda}^{\ast}$ (see Fig.~\ref{Sketch}(b)). (ii) This last step is necessary because this ``phase transition" is characterized by strong fluctuations of an order parameter field in the transition region \cite{Landau1980,Kadanoff1971}. These fluctuations are modeled here by a set of generalized velocity fields $\hat{v}_n$ given by
\begin{equation}
\hat{v}_{n} = L\langle(\partial_1 u_1)^n\rangle^{\frac{1}{n}} \,. 
\end{equation}
We can also define the generalized velocity based on the fluctuating acceleration field $\hat{a}_n$ given by
\begin{equation}
\hat{a}_{n}=L\langle(\partial_1 u_1)^{2n}\rangle^{\frac{1}{n}}\,.
\label{an}
\end{equation}
Points (i) and (ii) can now be combined, using \eqref{an} for the generalized velocity, to define an order-independent microscale Reynolds number
\begin{equation}
\hat{R}_{\lambda,n}= \sqrt{\frac{5}{3\langle\epsilon\rangle\nu}} L \hat{a}_n
=  \sqrt{\frac{5}{3\langle\epsilon\rangle\nu}} A_{2n}^{\frac{1}{n}} u_{\rm rms}^2 {\rm Re}^{\frac{\rho_{2n}}{n}}\,.
\end{equation}
Note that $L \hat{a}_n$ carries a physical dimension of length$^2$/time$^2$ for all $n$.
Taking $\beta_{\epsilon}=\langle\epsilon\rangle L/u_{\rm rms}^3$ as the dimensionless bulk energy dissipation rate, we get
\begin{equation}
\hat{R}_{\lambda,n}= \sqrt{\frac{5 A_{2n}^{\frac{2}{n}}}{3\beta_{\epsilon}}} {\rm Re}^{\frac{1}{2}+\frac{\rho_{2n}}{n}}\,.
\label{Re_rel}
\end{equation}
The driving of the isotropic flow, which is restricted to scales $r\approx L$, requires the further assumptions \cite{Yakhot2017} that the forcing is Gaussian and white in time, and injects turbulent kinetic energy in such a way that the mean kinetic energy dissipation rate is {\em independent} of the Reynolds number. The latter implies that $d_1=0$. In accordance with the collection of DNS results of decaying and forced turbulence in ref. \cite{Sreenivasan1998}, 
we can set $\beta_{\epsilon}\approx 0.4$ and thus $\sqrt{5/(3\beta_\epsilon)}\approx 1$.

The next part of the strategy is to calculate theoretically the unique value of the rescaled Reynolds number at which the transition takes place. We can then obtain, by matching the Gaussian behavior at the low Reynolds number with the power-law part with anomalous scaling (see Fig \ref{Sketch}b), the exponents $d_n$ and $\rho_{2n}$. We are in the fortunate position that the renormalization group theory \cite{Yakhot1986,Yakhot1992,Yakhot1992a} for HIT provides such a theory. We now take three specific steps: 

(a) We first establish a relation between $\rho_{2n}$ and $d_n$ by using arguments outlined in \cite{Yakhot2006,Schumacher2007}. In the limit of vanishing distances $r$ of a longitudinal velocity increment, the velocity is an analytic function such that the $x_1$-derivative of $u_1$ is defined as
\begin{align}
\frac{\partial u_1}{\partial x_1}\approx \frac{u_1(x_1+\eta)-u_1(x_1)}{\eta} 
                                               \equiv \frac{\Delta_{\eta}u_1}{\eta}\,.
\end{align}
The scale $\eta$ is a still-unknown fluctuating length scale distributed around the Kolmogorov dissipation length $\eta_K=\nu^{3/4}/\langle\epsilon\rangle^{1/4}$. Viscous effects become important when a local Reynolds number $Re_{\eta}$ is approximately unity, a property that is used in \cite{Paladin1987,Frisch1991,Yakhot2006}. Such a Reynolds number is given by
\begin{equation}
{\rm Re}_{\eta}=\frac{\eta \Delta_{\eta}u_1 }{\nu}\approx 1\,.
\label{methods1}
\end{equation}
Thus follows the relation $\Delta_{\eta}u_1 = \nu/\eta$, leading to the consequence that 
\begin{equation}
\partial_1 u_1\approx \frac{(\Delta_{\eta}u_1)^2}{\nu} \quad\mbox{and}\quad \epsilon\approx \frac{(\Delta_{\eta}u_1)^4}{\nu}\,.
\label{methods2}
\end{equation}
For the following, we assume that these relations are exact. Relations \eqref{methods2} are now used to rewrite \eqref{Mn} as
\begin{equation}
{\rm Re}^{\rho_{2n}}=\frac{L^{2n}}{A_{2n} u_{\rm rms}^{2n}} M_{2n} = \frac{1}{A_{2n}} {\rm Re}^{-2n} \left(\frac{L}{\eta}\right)^{4n}\,,
\label{methods3}
\end{equation}
and \eqref{En} as
\begin{equation}
{\rm Re}^{d_{n}}=\frac{L^{n}}{B_{n} u_{\rm rms}^{3n}} E_{n} =  \frac{1}{B_{n}} {\rm Re}^{-3n} \left(\frac{L}{\eta}\right)^{4n}\,.
\label{methods4}
\end{equation}
Consequently, the relation $B_n {\rm Re}^{d_{n}+n} = A_{2n} {\rm Re}^{\rho_{2n}}$ follows by comparing Eqs. \eqref{methods3} 
and \eqref{methods4} and implies
\begin{equation}
d_n+n = \rho_{2n} \quad \mbox{and}\quad B_n = A_{2n}\,.
\label{methods5}
\end{equation}

(b) Using the last relation in (\ref{methods5}), we rewrite \eqref{Re_rel} as follows:
\begin{equation}
\hat{R}_{\lambda,n} = \sqrt{\frac{5 A_{2n}^{\frac{2}{n}}}{3\beta_{\epsilon}}} {\rm Re}^{\frac{1}{2}+\frac{d_n+n}{n}} = 
B_{n}^{\frac{1}{n}} {\rm Re}^{\frac{1}{2}+\frac{d_n+n}{n}}\,,
\end{equation}
and thus, together with \eqref{methods5}, we have
\begin{equation}
{\rm Re}=\left[B_{n}^{-\frac{1}{n}} \hat{R}_{\lambda,n}\right]^{\frac{2n}{2d_n+3n}}=\tilde{B}_n \hat{R}_{\lambda,n}^{\frac{2n}{2d_n+3n}}\,.
\label{rel1}
\end{equation}   

(c) Finally, at the critical point of the phase transition to anomalous scaling, we have a unique Reynolds number for all $n$. That is, $R_{\lambda,n}^{\ast}=R_{\lambda}^{\ast}$ and ${\rm Re}^{\ast}={\rm Re}^{\ast}_n/C_n$ where $C_n$ is a slowly varying function of $n$. The slow variation of $C_n$ is supported by the DNS \cite{Yakhot2017}. Thus, Eq.\ \eqref{rel1} gives 
\begin{align}
{\rm Re}^{\ast}=\frac{{\rm Re}_n^{\ast}}{C_n}&=\tilde{B}_n (\hat{R}^{\ast}_{\lambda,n})^{\frac{2n}{2d_n+3n}} \nonumber\\ 
&\Rightarrow {\rm Re}_n^{\ast}
\approx C (\hat{R}^{\ast}_{\lambda,n})^{\frac{2n}{2d_n+3n}}\,.
\label{rel1a}
\end{align}   
In the last step, we use this weak $n$-dependence to simplify $C\approx C_n\tilde{B}_n$ for all $n$. For $n=1$, it follows that ${\rm Re}^{\ast}={\rm Re}^{\ast}_1=C (\hat{R}_{\lambda,1}^{\ast})^{2/3}$ and thus $C$ can be obtained. We are now able to derive the exponents $d_n$ by requiring that the turbulent and laminar Gaussian scaling laws to match at $R_{\lambda}^{\ast}=R_{\lambda,n}^{\ast}$. In detail, we obtain
\begin{equation}
(2n-1)!!= \left[C\left(\hat{R}^{\ast}_{\lambda,n}\right)^{\frac{2n}{2d_n+3n}} \right]^{d_n-n d_1}\,.
\label{matchrel}
\end{equation}     
In particular, the following three steps are used to solve the problem 
\begin{equation}
d_n=f(n, \hat{R}_{\lambda,n}^{\ast},C, d_1)\,.
\end{equation}     
(i) Use $d_1=0$ as a consequence of the applied forcing; (ii) $3\beta_{\epsilon}/5 =1$ as already discussed above; finally, also as stated earlier, 
(iii) the rescaled Taylor microscale Reynolds numbers are set to $\hat{R}^{\ast}_{\lambda}\equiv\hat{R}^{\ast}_{\lambda,n}$ for all $n$. 
The specific value of $\hat{R}^{\ast}_{\lambda}\approx 9$ follows from the renormalization group theory for the derivation of turbulence models 
\cite{Yakhot1992,Yakhot1992a,Yakhot2014}, supported by simulations in \cite{Schumacher2007}. Thus we are left with the relation
\begin{equation}
d_n=f(n)\,,
\end{equation}     
and the matching condition \eqref{matchrel} simplifies to
\begin{equation}
\log \left[2^n\frac{\Gamma(n+\frac{1}{2})}{\sqrt{\pi}} \right] = d_n \log C + \frac{2n d_n}{2d_n+3n} \log \hat{R}^{\ast}_{\lambda}\,,
\label{relmatch}
\end{equation}     
where we have used the relation between the double factorial and the Gamma function. This gives a quadratic equation for $d_n$ that can be solved for each order $n>1$ as done in Yakhot and Donzis \cite{Yakhot2017}. From this, one obtains $d_2=0.157$ and $d_3=0.489$. Similar predictions for the exponents $d_n$ can be obtained within the multifractal framework \cite{Paladin1987,Frisch1991,Nelkin1990,Biferale2008,Benzi2009}. For a recent application 
to Burgers turbulence we also refer to \cite{Friedrich2018}.

This completes the description of the theory used by Yakhot and Donzis \cite{Yakhot2017}. The theory is specific to HIT on at least two important counts: (1) the assumption of Gaussian white-in-time forcing and the use of the renormalization result that the transition Reynolds number is about 9. Clearly, the exponents $d_n$ are sensitive to both of these conditions. Yet, the theory introduces the ostensibly powerful concept that the scaling exponents in the turbulent state are entirely determined by the forcing and the transition Reynolds number. To claim any universality to these theoretical ideas, as Yakhot and Donzis intended, there has to be some concrete evidence from at least one more flow that does not belong to the HIT class. This is explored in the rest of the paper.

\section{Thermal convection model}
In convection, the buoyancy field is the product of the acceleration due to gravity, $g$, multiplied by a density contrast. It is given by 
\begin{equation}
B({\bm x},t)=-g\frac{\rho({\bm x},t)-\rho_0}{\rho_0}\,,
\end{equation}
where $\rho$ is the mass density field and $\rho_0$ a reference value. In a Boussinesq system with $\rho({\bm x},t)=\rho_0[1- \alpha (T({\bm x},t)-T_0)]$, the result is the well-known buoyancy term $g\alpha(T-T_0)$ that is added on the right hand side of the Navier-Stokes equation for the vertical velocity component $u_z$; $\alpha$ is the thermal expansion coefficient. The equations are made dimensionless by substituting space coordinates $x_i$, time $t$, velocity fields $u_i$, pressure field $p$, and temperature field $T$ by $\tilde{x}_i H$, $\tilde{t} H/U_f$, $\tilde{u}_i U_f$, $\tilde{p} \rho_0 U_f^2$, and $\tilde{T}\Delta T$, respectively. This implies that $\tilde{B}=\tilde{T}$. Here, $H$ is the height of the cell, $U_f=\sqrt{g\alpha\Delta T H}$ is the free-fall velocity, and $\Delta T>0$ is the temperature difference between the bottom and top plates.  
\begin{figure}
\centering
\includegraphics[scale = 0.8]{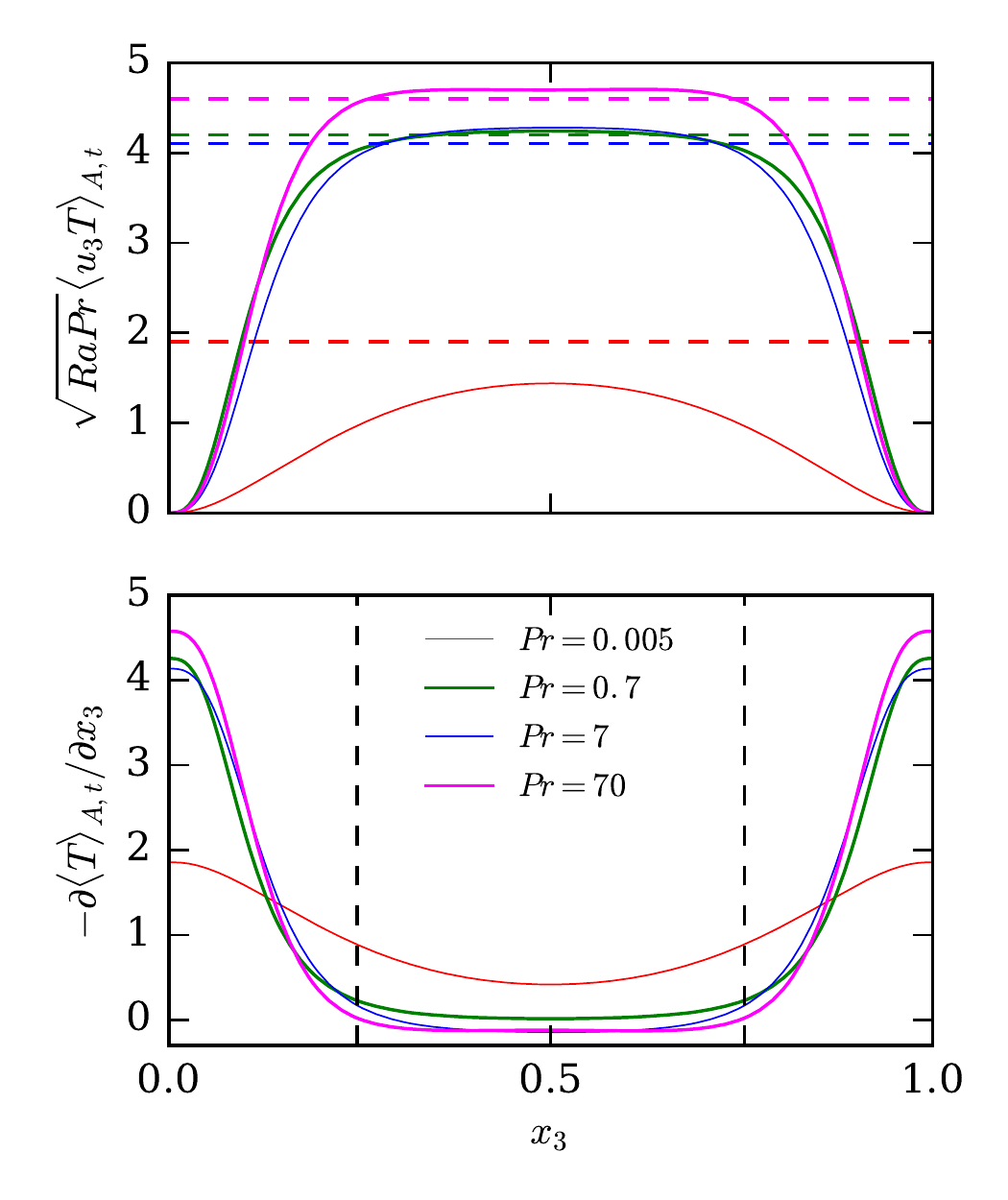}
\caption{Plane- and time-averaged vertical mean profiles of the convective heat current $j_{\rm conv}(x_3)$ (top) and conductive heat current $j_{\rm cond}(x_3)$ as defined in Eq.~\eqref{current}. Runs for ${\rm Pr}=0.005$, 0.7, 7, and 70 are displayed. The horizontal dashed line corresponds to Nusselt number Nu, as given in Table I. We also indicate the bottom and top end positions, $\ell_1$ and $\ell_2$ (see Table I), of the bulk volume that is used for statistical analysis.}
\label{fig1}
\end{figure}

We solve the coupled three-dimensional equations of motion for velocity field $u_i$ and temperature field $T$ in the Boussinesq approximation of thermal convection. They are given in dimensionless form by ($i,j=1,2,3$) 
\begin{align}
\label{ceq}
\frac{\partial \tilde{u}_i}{\partial \tilde{x}_i}&=0\,,\\
\label{nseq}
\frac{\partial  \tilde{u}_i}{\partial  \tilde{t}}+\tilde{u}_j \frac{\partial \tilde{u}_i}{\partial \tilde{x}_j}
&=-\frac{\partial \tilde{p}}{\partial \tilde{x}_i}+\sqrt{\frac{\rm Pr}{\rm Ra}} \frac{\partial^2 \tilde{u}_i}{\partial \tilde{x}_j^2}+  \tilde{B} \delta_{i3}\,,\\
\frac{\partial  \tilde{T}}{\partial  \tilde{t}}+\tilde{u}_j \frac{\partial \tilde{T}}{\partial \tilde{x}_j}
&=\frac{1}{\sqrt{{\rm Ra} {\rm Pr}}} \frac{\partial^2 \tilde{T}}{\partial \tilde{x}_j^2}\,.
\label{pseq}
\end{align}
Here the Rayleigh number ${\rm Ra}=g\alpha\Delta T H^3/(\nu\kappa)$. The aspect ratio of the cell is $\Gamma = L/H=25$, with the cross-section of the cell being $L \times L$.
No-slip boundary conditions for the fluid are applied at all walls. The top and bottom plates are held at constant dimensionless temperatures $\tilde{T}=0$ and 1, respectively. The side walls are thermally insulated. The equations are numerically solved by the Nek5000 spectral element method package \cite{nek5000} which converges exponentially fast and resolves the velocity derivatives accurately \cite{Scheel2013,Pandey2018}. Table 1 summarizes all the runs analyzed and lists a few important parameters. From now on, for simplicity, we will drop the tilde for dimensionless quantities. 

The turbulent heat transfer can be decomposed into two contributions that sum up to a constant, a conductive and a convective heat current. In dimensionless form they can be written as
\begin{align}
\label{current}
{\rm Nu} &=j_{\rm conv}+j_{\rm cond}\nonumber\\
      &=\sqrt{{\rm Ra} {\rm Pr}} \langle u_3 T(x_3)\rangle_{A,t}-\frac{\partial\langle T(x_3)\rangle}{\partial x_3}\,, 
\end{align}
where Nu denotes the Nusselt number. Figure \ref{fig1} displays mean vertical profiles of both currents which are obtained by averages with respect to the horizontal planes $A$ and time $t$. It is seen that the Nusselt number is significantly reduced for the low Prandtl number case. Note also that ${\rm Nu}(x_3)$ = constant for all the cases discussed. For ${\rm Pr}\ge 0.7$, the magnitude of $Nu$ is smaller than the convective heat flux (see top panel of Fig. \ref{fig1}). This is in line with a finite {\em positive} slope of the mean temperature profile $\langle T(x_3)\rangle_{A,t}$ in the bulk, as is visible in the bottom panel of Fig \ref{fig1}. 
\begin{table*}
\begin{ruledtabular}
\begin{center}
\begin{tabular}{lcccccccc}
              &  ${\rm Pr}$    & $N_e$  &   $N$  &  ${\rm Nu}$  & $\ell_1$ & $\ell_2$ &  ${\rm Re}_{\ell}$  &  $\;\;\langle\epsilon\rangle_{V_{\ell}}\;\;$\\
\hline
Run 1$^{\ast}$       &  100      &  1,352,000  &  5   & $4.6 \pm 0.003$ &  0.247  & 0.753 & 0.44  & $2.2\times 10^{-4}$  \\                      
Run 2       &  70        &  1,352,000  &  5   & $4.6 \pm 0.01$  &  0.247  & 0.753  & 0.63  & $4.0\times 10^{-4}$  \\                      
Run 3       &  35        &  1,352,000  &  5   & $4.5 \pm 0.01$  &  0.247  & 0.753  & 1.23  & $6.6\times 10^{-4}$  \\
Run 4       &   7         &  1,352,000  &  5   & $4.1 \pm 0.01$  &  0.247  & 0.753  & 5.58  & $2.1\times 10^{-3}$  \\
Run 5       &  0.7       &  1,352,000  &  5   & $4.2 \pm 0.02$  &  0.247  & 0.753  & 48.9  & $7.8\times 10^{-3}$  \\                      
Run 6$^{\ast}$       &  0.3       &  1,352,000  &  5   & $4.0 \pm 0.01$  &  0.247  & 0.753  & 96.7  & $1.2\times 10^{-2}$  \\
Run 7$^{\ast}$       &  0.1       &  1,352,000  &  7   & $3.5 \pm 0.01$  &  0.247  & 0.753  & 215   & $1.9\times 10^{-2}$  \\
Run 8       &  0.021   &  2,367,488  &  7   & $2.6 \pm 0.01$  &  0.223  & 0.777  & 636   & $2.9\times 10^{-2}$  \\
Run 9       &  0.005   &  2,367,488  &  11 & $1.9 \pm 0.01$  &  0.223  & 0.777  & 1408 & $3.3\times 10^{-2}$  \\
\end{tabular} 
\caption{Parameters of the different direct numerical simulations. The aspect ratio is always $\Gamma=25$, the Rayleigh number is always ${\rm Ra}=10^5$. The Prandtl number, Pr, the number of spectral elements, $N_e$, the polynomial order of the expansion of all fields on each element in each spatial direction, $N$, the Nusselt number, Nu, the lower and upper heights of the bulk region analyzed,
$\ell_1$ and $\ell_2$, and the resulting Reynolds number ${\rm Re}_{\ell}$ with $\ell=\ell_2-\ell_1$ (see Eq.~(\ref{Reell})) are listed. The last column contains the mean value of the kinetic energy dissipation rate in the bulk volume $V_{\ell}$. The runs which are indicated with an asterisk are conducted in addition to those of \cite{Pandey2018}.}
\end{center}
\end{ruledtabular}
\end{table*}
\begin{figure}
\centering
\includegraphics[scale = 0.7]{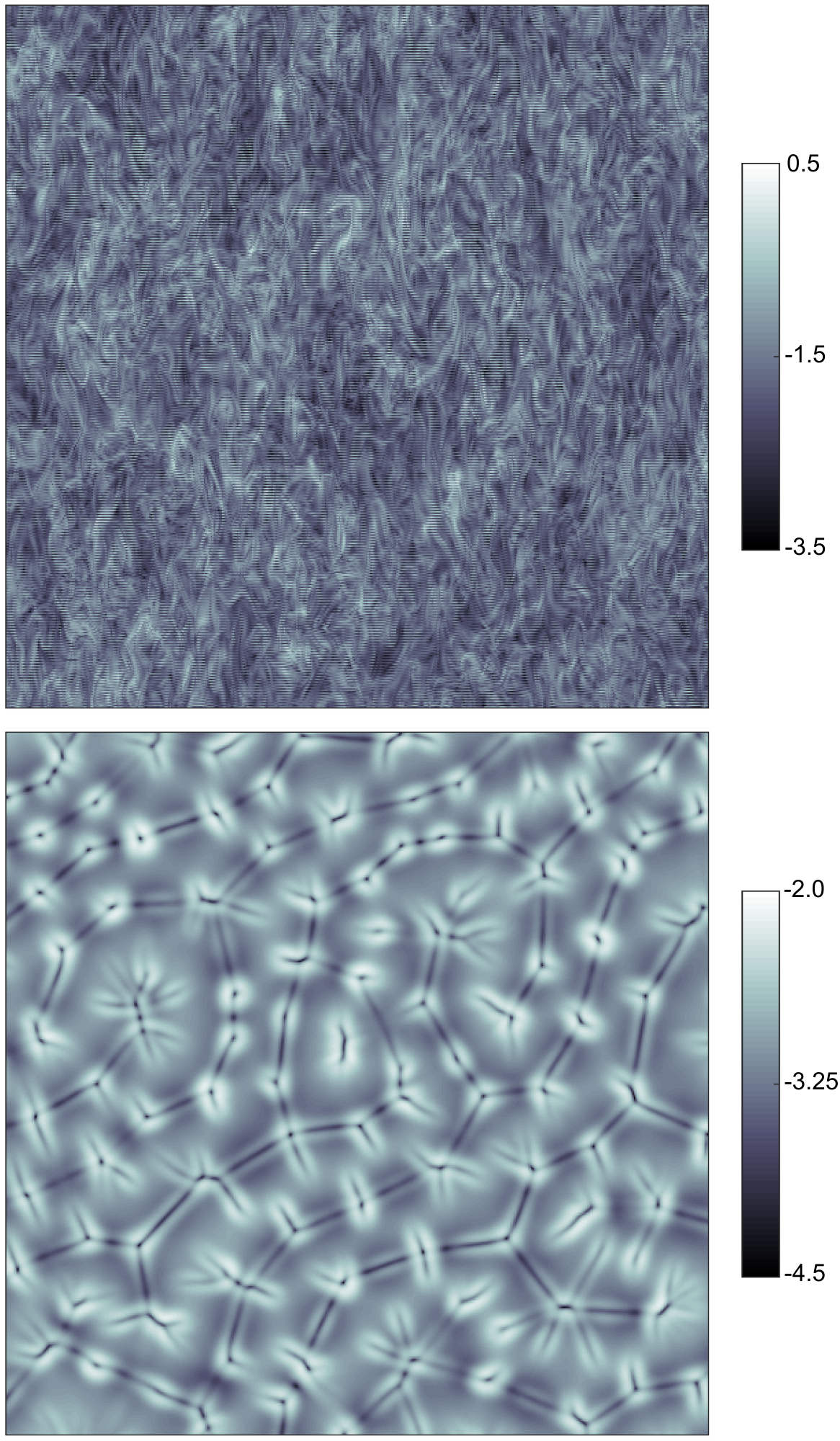}
\caption{Contour plots of the instantaneous kinetic energy dissipation rate $\epsilon({\bm x},t)$ at the mid-plane, $z=1/2$. Contour levels are displayed 
in units of the decadal logarithm. Top: ${\rm Pr}=0.005$. Bottom: ${\rm Pr}=70$. Only a quarter of the full cross section is shown.}
\label{fig2}
\end{figure}

\section{Statistical analysis}
\subsection{Normalized energy dissipation rate}
We consider here only the energy dissipation to make our main point; the velocity derivatives as well as the vorticity have been computed and the conclusions drawn from their behavior are similar. Figure \ref{fig2} displays contour plots of mid-plane cross-sections of the instantaneous kinetic energy dissipation rate field. The levels are given in units of the decadal logarithm. We display snapshots for the two runs at the smallest (top) and one of the largest (bottom) Prandtl numbers. The differences in the fine structure of the two fields is evident. Low-Pr convection is known to be highly inertial \cite{Pandey2018,Schumacher2015}, as can be seen here clearly.   
 
The statistical analysis to be discussed below is always restricted to the fraction of the convection layer between heights $\ell_1$ and $\ell_2$ highlighted roughly by vertical lines in Fig. \ref{fig1}; the exact values are listed in Table I. The amplitude of the mean kinetic energy dissipation rate in this region varies systematically with Pr and thus with ${\rm Re}_\ell$, as indicated in the Table. This Reynolds number, which corresponds to Re in the HIT case, 
is given by
\begin{equation}
{\rm Re}_{\ell} = \frac{u^{(\ell)}_{\rm rms}\ell}{\nu}=\sqrt{\frac{\rm Ra}{\rm Pr}} \, \ell \sqrt{\langle u_x^2+ u_y^2+ u_z^2\rangle_{V_{\ell}}} \,, 
\label{Reell}
\end{equation}
where $\ell$ is the thickness of the bulk region (which is outside the thermal boundary layers), 
$V_{\ell}=A\ell$ and $A = L\times L$ being the cross sectional area of the cuboid cell (see again Table I).
Since we are interested in the small-scale fluctuations, we decompose the velocity and temperature fields as follows
\begin{align}
{\bm u}^{\prime}({\bm x},t) &= {\bm u}({\bm x},t)-\langle {\bm u}\rangle_t({\bm x})\,,\nonumber\\
T^{\prime}({\bm x},t) &= T ({\bm x},t)-\langle T\rangle_t({\bm x})\,,\nonumber
\end{align}
In dimensionless form the kinetic energy dissipation rate is then given by 
\begin{equation}
\epsilon({\bm x},t) =\frac{1}{2}\sqrt{\frac{Pr}{Ra}} \left({\bm\nabla {\bm u}^{\prime}}+({\bm\nabla {\bm u}^{\prime}})^T\right)^2\,. 
\end{equation}
See also Eq.~(\ref{ediss}) for comparison.  

\begin{figure*}
\centering
\includegraphics[scale = 0.8]{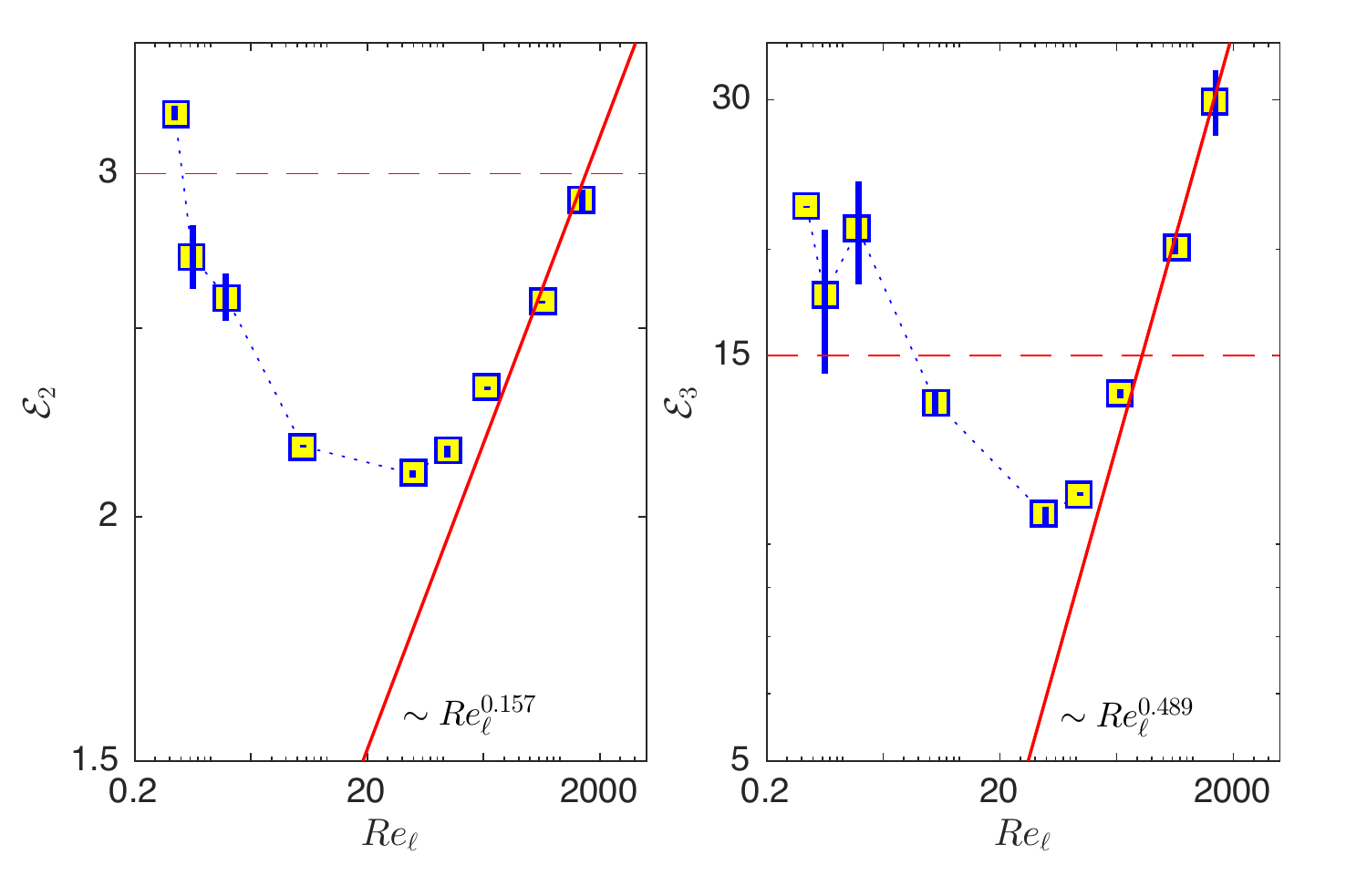}
\caption{Normalized moments ${\cal E}_n$ of kinetic energy dissipation rate versus Reynolds number ${\rm Re}_{\ell}$ for orders $n=2,3$. We have verified that these two moments converge. We have not shown data for $n>3$ because their convergence is doubtful. The solid line is a power law scaling of the theory \cite{Yakhot2006}. The horizontal dashed lines indicate moments in correspondence with a Gaussian distribution, ${\cal E}_n=(2n-1)!!$. Error bars have been obtained as the difference of the normalized moments when calculated for the first and second halves of the corresponding data record.}
\label{fig3}
\end{figure*}

The data for the normalized moments ${\cal E}_n({\rm Re}_{\ell})$ for orders $n=2$ and $n=3$ are summarized in Fig. \ref{fig3}. These moments at high Reynolds numbers indeed follow the expected scaling laws \cite{Yakhot2006,Schumacher2007,Yakhot2017}. The transition Reynolds number ${\rm Re}_{\ell}\approx 100$ -- 200 also corresponds well with the value reported for homogeneous and isotropic turbulence. These two features are in accord with a universal transition and subsequent universal scaling. However, the major difference from the schematic in Fig. 1 is that the Reynolds number dependence in the pre-transition region is non-monotonic. The data at the lowest Reynolds numbers are indeed roughly comparable to $(2n-1)!!$, as indicated by the horizontal lines, but pass through a minimum before following the expected power-laws. In the rest of this section, we will consider the low-Reynolds-number behavior and how, if at all, the non-monotonic behavior of the data may still be consistent with the spirit of the theory of section 2. 
  
At very low Reynolds numbers prior to the onset of rising and falling thermal plumes, it is conceivable that the flow starts with a nearly Gaussian forcing, with dissipation moments given by $(2n-1)!!$. However, as the Reynolds number increases the small-scale fluctuations are mostly determined by the plumes. This is a significant difference from the low-Reynolds-number flows in \cite{Yakhot2017}, which are always driven by {\em stochastic} forces. For convection, the momentum balance of the Boussinesq equations requires that
\begin{equation}
g\alpha T^{\prime} \sim \frac{\partial u_z^{\prime}}{\partial t}\sim W_{\rm pl} \frac{\partial u_z^{\prime}}{\partial z} \sim 
\sqrt{f_{\rm pl}\,\epsilon}\,,
\label{estimate}
\end{equation}
where $W_{\rm pl}$ and $f_{\rm pl}$ are typical rising velocities and thermal plume detachment frequencies, respectively. They have been discussed, for example in \cite{Castaing1989}. This relation would imply that the statistics of the kinetic energy dissipation rate are connected to those of the temperature fluctuations, and so we shall discuss the nature of temperature fluctuations next.   

\begin{figure*}
\centering
\includegraphics[scale = 0.6]{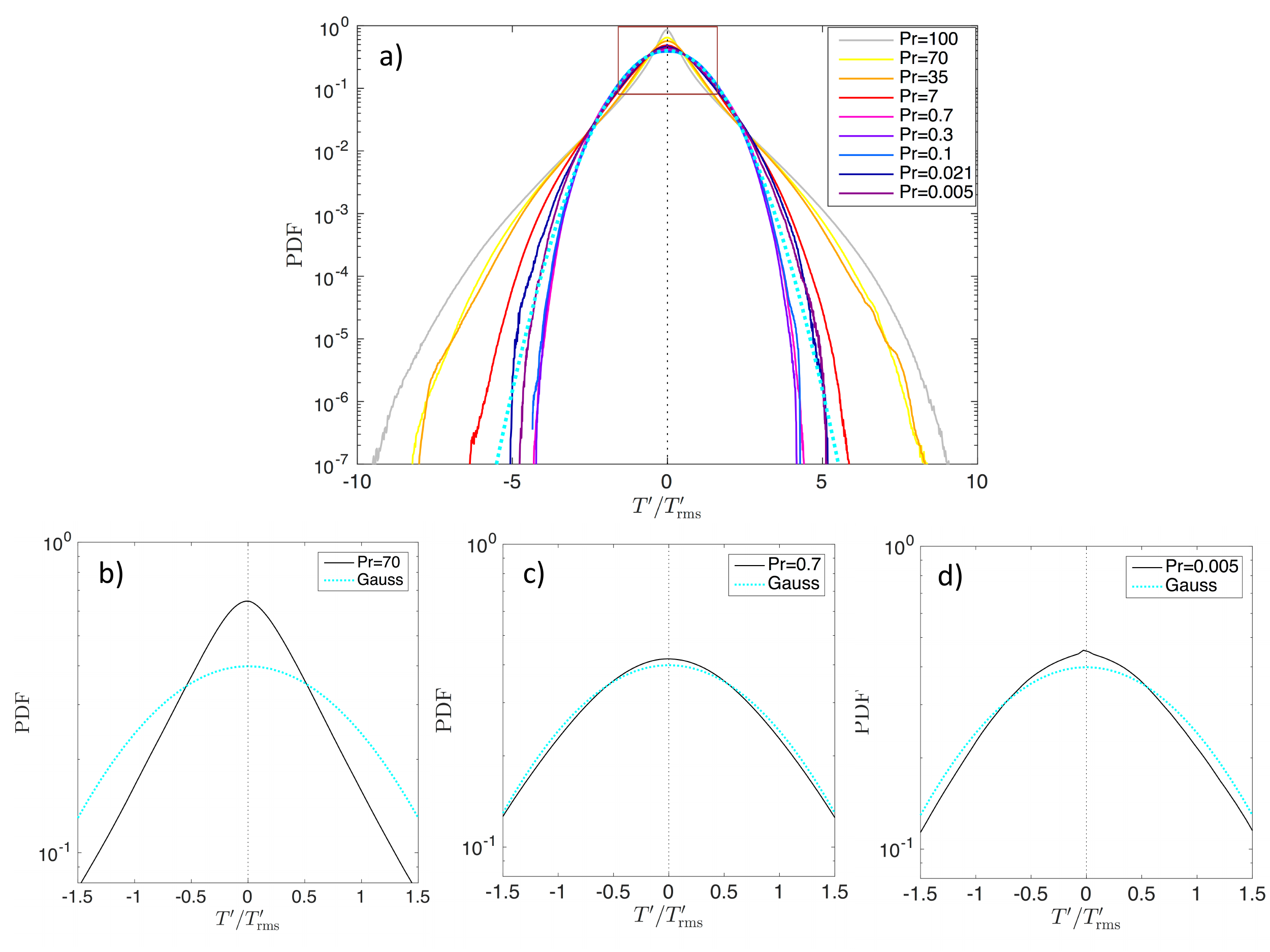}
\caption{Probability density functions of the temperature fluctuations in the bulk of the convection layer. Temperature amplitudes are normalized by the corresponding root-mean-square value. The Gaussian distribution is added as a dashed line for comparison. (a) All data sets are shown. The small box 
at the top center of panel (a) indicates the range that is shown in panels (b) to (d) where selected data sets are replotted. (b) Pr=70. (c) Pr=0.7. (d) Pr=0.005.}
\label{fig5}
\end{figure*}
 
\subsection{Temperature fluctuations}
The PDFs of the temperature fluctuations are obtained in the same bulk volume $V_{\ell}$ as energy dissipation. Figure \ref{fig5} (a) plots all data together with a Gaussian PDF (dashed line). The data at the highest Prandtl numbers develop the fattest tails while the remaining runs for ${\rm Pr}\le 0.7$ depart only slightly from Gaussian. 
\begin{figure}
\centering
\includegraphics[scale = 0.45]{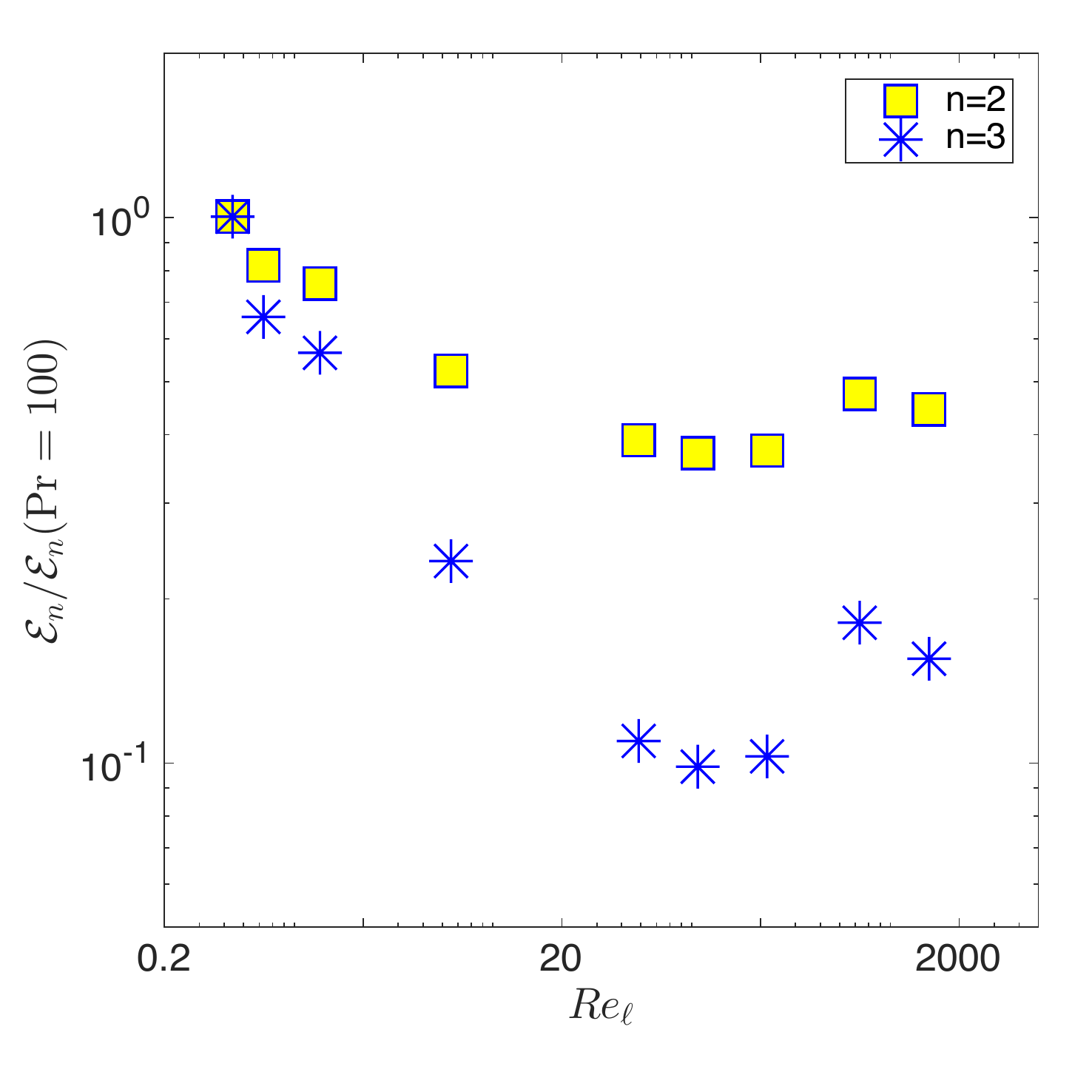}
\caption{Rescaled energy dissipation moments versus Reynolds number ${\rm Re}_{\ell}$ for orders $n=2$ and 3. These data are obtained from 
the probability density functions of the temperature fluctuations in Fig. \ref{fig5}.}
\label{fig6}
\end{figure}

Predictions for the shape of the temperature PDF in convection have been worked out in \cite{Yakhot1989,Yakhot1990}. According to this work, Gaussian temperature distributions follow when no particular velocity scale is present in the local convective heat flux $u_3^{\prime}T^{\prime}$, which is the production term for turbulent kinetic energy. An exponential distribution occurs when a characteristic plume velocity exists. Both functional forms were derived in \cite{Yakhot1990} for small values of the argument, $X=T^{\prime}/T^{\prime}_{\rm rms} \lesssim 1$ and thus not related to the tails of the PDF of the temperature fluctuations. Therefore, our obtained PDFs are magnified and replotted in Fig. \ref{fig5} (b--d) for $|X|<1.5$ for three out of the nine data sets. It is clear from this plot that the PDFs of the temperature fluctuations for the lowest Reynolds (or highest Prandtl) numbers behave more like an exponential distribution than a Gaussian one (see panel (b) of Fig. \ref{fig5} for $Pr=70$). In contrast, the PDFs of temperature fluctuations for higher Reynolds (or lower Prandtl) numbers are close to Gaussian in the center with sub-Gaussian tails, as seen in panels (c) and (d) of the same figure. 

Our argument based on \eqref{estimate} is supported by Fig. \ref{fig6} where we replot the dissipation rate moments as a function of the Reynolds number.  The data are moments based on the PDFs of the temperature fluctuations via the substitution $T^{\prime}\sim \sqrt{\epsilon}$ from (\ref{estimate}). The same {\em qualitative} crossover behavior as the original data in Fig. \ref{fig3} is observable. Since no quantitative estimate can be made, we took the lowest Reynolds number data as a reference in Fig. \ref{fig6}.

Again, for the intermediate Reynolds number regime between the Gaussian state and the turbulent state, a major change occurs which renders the moments of the energy dissipation lower than $(2n-1)!!$. We may speculate, for instance, that the forcing is then generated by stronger plumes which are still infrequent enough for them not to merge; this might push the moment values to lower numbers leading to the observed minimum that seems to come close to 
exponential statistics, ${\cal E}_n\sim n!$. We may thus enlarge the theoretical construct of section 2 in the following manner. A flow might always start at the lowest Reynolds number with Gaussian forcing but, in natural flows like convection, one may develop an intermediate state in which the driving is no longer Gaussian and white in time. This state usually precedes the turbulent state, which makes the transition process non-universal, though the turbulent state may well be universal.

\section{Summary and discussion}

In refs.\ \cite{Yakhot2006,Yakhot2017}, a theory was developed to understand self-consistently the evolution of homogeneous and isotropic turbulence subject to a Gaussian forcing that is white in time. The flow was numerically shown to evolve from a state in which the moments of energy dissipation proceeded from $(2n-1)!!$ at low Reynolds numbers through a known transition point to become turbulent with anomalous scaling exponents. The transition point was known in the sense that it was computed by a renormalization group approach to turbulence modeling \cite{Yakhot1992,Yakhot1992a}. Matching at this transition point the Gaussian initial state and the anomalous turbulent state yielded the scaling exponents in the latter. This led to the speculation that anomalous exponents in the turbulent state were determined entirely by the low-Reynolds-number state of the flow and the transition point. This is indeed a powerful conclusion if true, and can be advanced only by subjecting it to further tests. This has been the purpose of the paper.

After restating the theory to clarify its assumptions, we examined the data in recent convection simulations \cite{Pandey2018}. The low-Reynolds-number regime consists of two branches. We found that the flow at the lowest Reynolds numbers behaves as if the forcing is Gaussian which is indicated by 
${\cal E}_2 \sim 3!!$ and ${\cal E}_3\sim 5!!$. It is followed by a regime that loosely resembles exponential statistics. The transition to the anomalous
scaling proceeds for ${\rm Re}_{\ell}\sim 10^2$  which is interestingly at the same order of magnitude as that found in \cite{Yakhot2017}. 
The anomalous scaling exponents 
$d_n$ are the same as in the flow with Gaussian white-in-time forcing. However, the most important difference is that the flow does not go directly from the initial state with Gaussian-like characteristics to the final turbulent state. We expect this last conclusion to be a general feature of transitional flows, with each flow developing its own (i.e., non-universal) intermediate state. This brings us to the conclusion that one needs to temper the notion that the initial state fully determines the turbulent state and its anomalous scaling exponents. Nevertheless, it appears fruitful to regard the Yakhot-Donzis theory as basic in some sense, and examine it further for putting it on a firmer basis.

The variation of the Reynolds number results from a variation of the Prandtl number at a fixed Rayleigh number in the present simulation data
record. This causes very different thicknesses of the viscous and thermal boundary layers with respect to each other and alters the structure of the thermal plumes, such as their stem width. As a part of the future work, we plan to conduct a series at ${\rm Pr}\equiv 1$ where an increase in Rayleigh number 
generates larger Reynolds numbers and to compare these results with the present findings.

\section*{Acknowledgements.} 
AP acknowledges support by the Deutsche For\-schungs\-gemeinschaft within the Priority Programme on Turbulent Superstructures under Grant No. DFG-SPP 1881. JS wishes to thank the Tandon School of Engineering at New York University for financial support. 
Computing resources at the Leibniz Rechenzentrum Garching are provided by the Large Scale Project with Grant No. pr62se of the Gauss Centre for Supercomputing.

\end{document}